\documentclass[twocolumn,aps,superscriptaddress]{revtex4}
\usepackage{graphicx}
\usepackage{float}
\usepackage{dcolumn}
\usepackage{bm}
\usepackage{color}
\usepackage{amsmath}
\usepackage{amssymb}
\usepackage{amsfonts}
\usepackage{esint}
\usepackage{times}
\usepackage{xcolor}
\usepackage{phaistos}
\usepackage{braket}
\usepackage{comment}
\usepackage{multirow}

\usepackage{pdfpages}

\usepackage[colorlinks,linkcolor=blue,anchorcolor=blue,urlcolor=blue,citecolor=blue]{hyperref}

\begin{document}

\title{A multiplexed control architecture for superconducting qubits \\ with row-column addressing}

\author{Peng Zhao}
\email{shangniguo@sina.com}
%\email{shangniguo@sina.com; Present address: Hefei National Laboratory, Hefei 230088, China}
\affiliation{Tongling, Anhui 244000, China}
\affiliation{Beijing Academy of Quantum Information Sciences, Beijing 100193, China}

\date{\today}

\begin{abstract}
In state-of-the-art superconducting quantum processors, each qubit is controlled by at least one control line
that delivers control pulses generated at room temperature to qubits operating at millikelvin temperatures. While this strategy has been
successfully applied to control hundreds of qubits, it is unlikely to be scalable to control thousands of qubits, let alone
millions or even billions of qubits needed in fault-tolerance quantum computing. The primary obstacle lies in the wiring
challenge, wherein the number of accommodated control lines is limited by factors, such as the cooling power, physical
space of the cryogenic system, the control footprint area at the qubit chip level, and so on. Here, we introduce a
multiplexed control architecture for superconducting qubits with two types of shared
control lines, row and column lines, providing an efficient approach for parallel controlling $N$ qubits with $O(\sqrt{N})$
control lines. With the combination of the two-type shared lines, unique pairs of control pulses are delivered to qubits at
each row-column intersection, enabling parallel qubit addressing. Of particular concern here is that, unlike
traditional gate schemes, both single- and two-qubit gates are implemented with pairs of control pulses. Considering the
inherent parallelism and the control limitations, the integration of the architecture into quantum computing systems
should be tailored as much as possible to the specific properties of the quantum circuits to be executed. As such, the
architecture could be scalable for executing structured quantum circuits, such as quantum error correction circuits.

\end{abstract}

\maketitle

%far red-detuned
%%%%%%%%%%%%%%%%%%%%%%%%%%%%%%%%%%%%%%%%%%%%%%%%%%%%%%%%%%%%%%%%%%%%%%%%%%%%%

\section{Introduction}\label{SecI}

To perform quantum computing with noisy qubits for solving valuable problems that are intractable for classical
computing, quantum error correction (QEC) is widely recognized as the ultimate solution~\cite{Terhal2015,Dennis2002}. To date, among the various
candidate quantum systems for realizing QEC, superconducting qubits have been demonstrated as a leading one~\cite{Campbell2024,Krinner2022,Zhao2022,Sundaresan2023,Acharya2023a}. In a
superconducting quantum processor with tens of qubits, a recent experiment has shown that QEC begins to suppress
logical errors with increasing system sizes~\cite{Acharya2023a}. Nevertheless, to fully utilize the power of
QEC, such as surface code~\cite{Dennis2002,Fowler2012}, in quantum computing, it is generally believed that
high-fidelity control over millions or even billions of qubits is required~\cite{Fowler2012,Gidney2021}.

\begin{figure}[tbp]
\begin{center}
\includegraphics[keepaspectratio=true,width=\columnwidth]{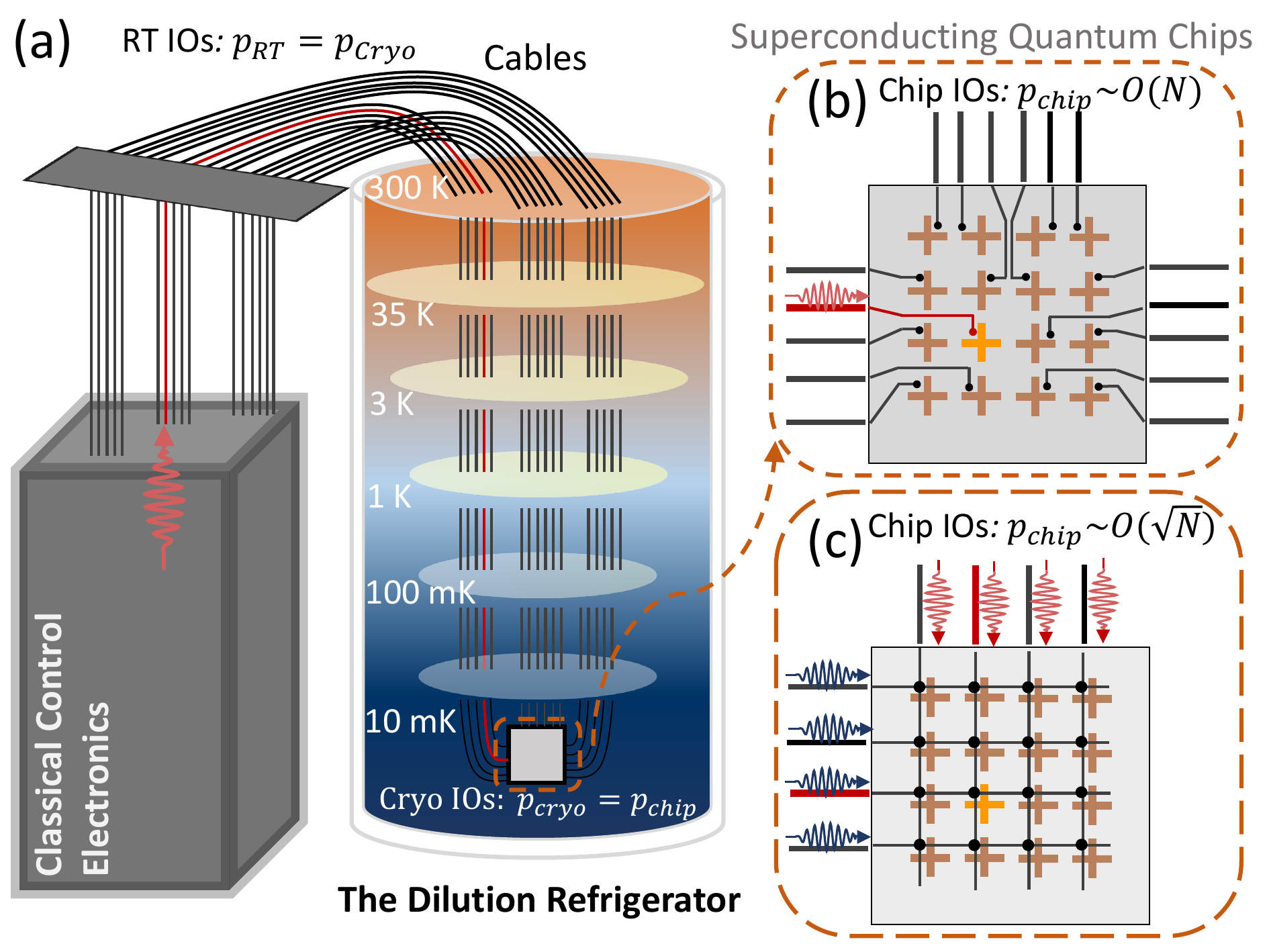}
\end{center}
\caption{Schematic (not to scale) of a typical superconducting quantum computing system. (a) Qubits
at millikelvin temperatures are controlled using control pulses
generated by classical electronics at room temperature and delivered through cables. The number of control
lines (input-output connections, IOs) at room temperature $P_{RT}$ and the number of cables in the
dilution refrigerator $P_{cryo}$ are determined by the number of input-output (IO) terminals
of the chip $P_{chip}$, leading to $P_{RT}=P_{cryo}=P_{chip}$. (b) With the independent
control, $P_{chip}\sim O(N)$, where $N$ is the qubit number. This strategy offers maximum flexibility
in qubit control, enabling individual addressing without affecting other qubits (see the qubit highlighted
in light orange). (c) In the multiplexed control architecture, applying appropriate
pulses to control lines simultaneously allows unique pairs of control pulses to be
delivered to qubits at the intersection of row lines and column lines, enabling
parallel qubit addressing. Hence, the IO terminals $P_{chip}$ scales with $\sqrt{N}$. Here, each
qubit (see the qubit highlighted in light orange) or any subgroup of qubits cannot be individually
addressed without affecting all other idle qubits.}
\label{fig1}
\end{figure}

In state-of-the-art superconducting quantum processors, shown in Fig.~\ref{fig1}(a), each qubit is
controlled by at least one control line, which delivers dedicated control pulses generated at room temperature to qubits
operating at millikelvin temperatures~\cite{Chen2018,Krinner2019,Krantz2019}. For controlling a $N$-qubit system,
the number of control lines (input-output connections, IOs) at room temperature ($P_{RT}$) and in the dilution
refrigerator ($P_{cryo}$) is dictated by the IO terminals of the qubit chip ($P_{chip}$), scaling linearly
with the qubit number $N$, see Fig.~\ref{fig1}(b), resulting in $P_{RT}=P_{cryo}=P_{chip}\sim O(N)$~\cite{Frankea2019}. This
independent control strategy offers great flexibility in controlling qubits~\cite{Krantz2019} and is applied to achieve low gate errors across
small-scale systems comprising tens or hundreds of qubits. However, the scalability of this approach to larger systems is hindered
by challenges such as heating loads from the control lines~\cite{Krinner2019}, non-negligible feature sizes of cables or IO
terminals, and the requirement of at least one digital-to-analog converter (DAC) per line for qubit control~\cite{Krantz2019}. These
limitations highlight the wiring challenge associated with scaling up quantum
processors~\cite{Vandersypen2017,Frankea2019,Reilly2019,Martinis2020}, where the number of manageable
control lines is limited by factors including cooling power, geometric constraints of the cryogenic system, control footprint area at
the qubit chip level, and the overhead of classical electronics. To facilitate a more intuitive understanding of this challenge, consider
that to control a superconducting quantum processor with a few thousand qubits, the number of required control lines or IOs would be comparable to that needed to manage one billion transistors in state-of-the-art classical processors~\cite{Vandersypen2017}. In this
sense, the wiring challenges need to be addressed before a large-scale superconducting quantum processor becomes feasible.

To alleviate the wiring challenge in superconducting quantum systems, various strategies which aim to reduce the control
lines running from room temperature to cryogenic temperature, i.e., achieving $P_{RT}<P_{Cryo}$, have been explored \cite{Asaad2016,Lecocq2021,McDermott2018,Bardin2019,Dijk2020,Chakraborty2022,Takeuchi2023a,Takeuchi2023b,Acharya2023}.
One of the most widely studied approaches is the implementation of cryogenic control electronics operated at 4 K~\cite{McDermott2018,Bardin2019,Dijk2020,Chakraborty2022,Takeuchi2023a,Takeuchi2023b} or 10 mK~\cite{Takeuchi2023a,Takeuchi2023b,Acharya2023},
but implementing successful qubit control at scale while achieving ultra-low-power dissipation remains challenging. Additionally, on-chip control
electronics have also been explored to reduce the chip IO terminals~\cite{McDermott2018,Liu2023}. However, as
the control footprint area at the chip level is limited by qubit size, besides the heating dissipation and the
newly added noises, the on-chip integration can increase the complexity of wire routing, especially when scaling up.
Therefore, a common view is that only taking these strategies alone is unlikely to address the
wiring challenge~\cite{Vandersypen2017}.

Besides the above top-down approaches, one alternative that has emerged from taking a bottom-up perspective
involves leveraging multiplexed qubit control with shared lines to reduce line overhead. This perspective has been extensively
explored for semiconductor spin qubits~\cite{Vandersypen2017,Hill2015,Veldhorst2017,Li2018} and superconducting quantum
annealing processors~\cite{Johnson2010,Bunyk2014}. For superconducting qubits, the most successful
demonstration is the frequency-multiplexed qubit readout~\cite{Jerger2011,Jerger2012,Chen2012}, where a single feedline is shared
by several readout circuitries, enabling simultaneous readout of several qubits. However, exploring the multiplexing for
qubit control is still in its infancy~\cite{Li2021,Bejanin2022,Zhao2023,Shi2023}. Unlike the independent qubit
control, the multiplexed control will degrade the control flexibility. Thus, to realize quantum
computing, generally, new control overhead or hardware components shall be introduced~\cite{Zhao2023,Shi2023}.
Most importantly, further demonstrations of these new features that are compatible with high-fidelity
qubit control are required.

In this work, we propose a multiplexed control architecture for superconducting qubits with row-column addressing,
offering the potential for parallel control of $N$ qubits with $O(\sqrt{N})$ control lines. By incorporating both row
and column shared control lines, unique pairs of control pulses are delivered
to qubits or couplers at each row-column intersection, allowing parallel single- or two-qubit gate
operations in qubit lattices.  Accordingly, we present various single- and two-qubit schemes that are both
compatible with the row-column addressing and the existing superconducting qubit technologies (with no new hardware components).
We further show that while the control flexibility is compromised, the inherent parallelism of the architecture makes it
well-suited for executing structured quantum circuits, which comprise layers of parallel single- or two-qubit gates across
qubit lattices, such as QEC circuits.

We note that very recently, a similar scheme for selectively addressing qubits has also
been studied in semiconductor spin qubits~\cite{Gyorgy2022,Bosco2023,John2023} by using two microwave drives and has been
proposed to enable parallel single-qubit gates~\cite{Gyorgy2022}.

The paper is organized as follows. In Sec.~\ref{SecII}, we first provide an overview of the multiplexed
control architecture. Then, in QEC with surface code, we illustrate that the integration of the architecture into
quantum computing systems should be tailored to specific quantum circuits. In Sec.~\ref{SecIII}, we
give detailed descriptions of the single- and two-qubit gate schemes that support row-column addressing. In Sec.~\ref{SecIV}, we
discuss the challenges to be faced when scaling up. Finally, in Sec.~\ref{SecV}, we provide
a summary of our study.

\section{The multiplexed control architecture with row-column addressing }\label{SecII}

\begin{figure*}[tbp]
\begin{center}
\includegraphics[width=16cm,height=6cm]{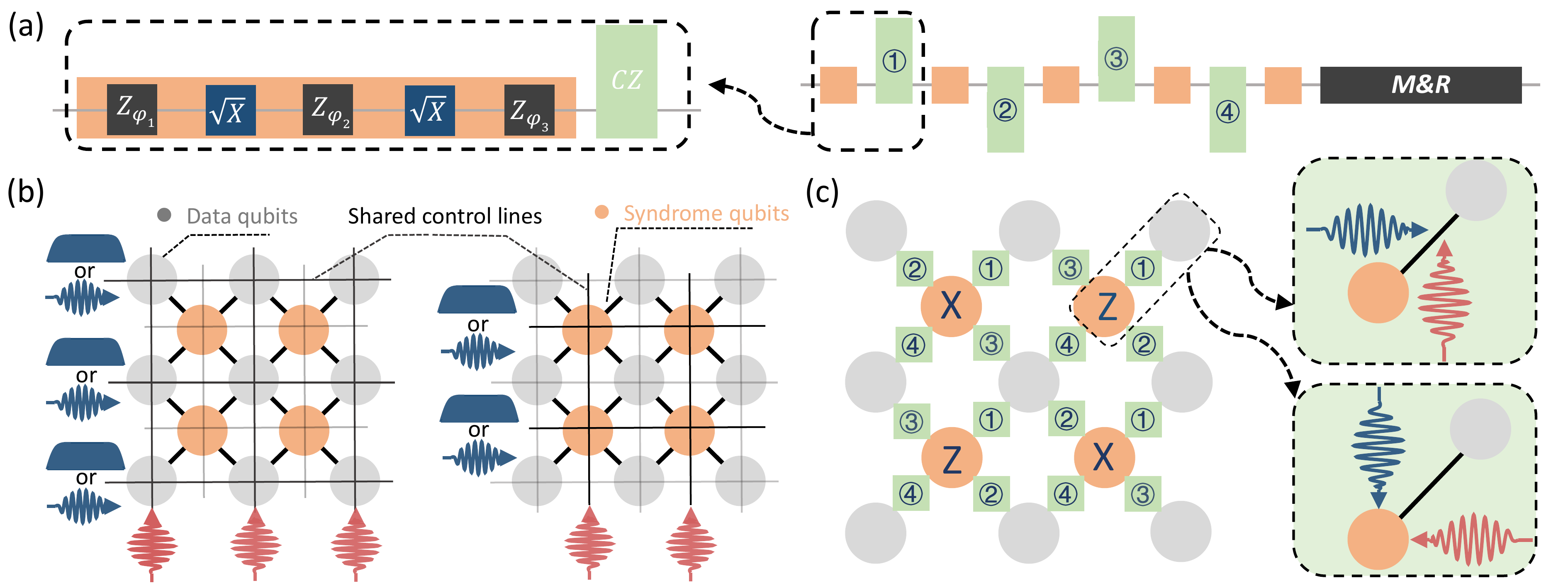}
\end{center}
\caption{Multiplexed qubit control for QEC with surface code. (a) A typical stabilizer measurement circuit in surface code
using $\textmd{CZ}$ gates, with the single-qubit gate decomposed into two $\sqrt{\textmd{X}}$ gates
and three $\textmd{Z}$ gates. During the QEC cycle, interleaved single- and two-qubit gates are implemented and at
the end of the cycle, syndrome qubits are measured first and then reset while data qubits are idle. (b) Two sets of
shared control lines are employed for independent addressing of data qubits and syndrome qubits. This setup ensures
compatibility between the multiplexed control architecture and QEC circuits, which involve dynamical decoupling sequences
to reduce qubit idling errors and two-qubit gates with one-qubit driving. The shared control
lines can deliver microwave or flux pulses for qubit control. (c) In QEC with surface code, the
sequence of two-qubit gates depends on the type of syndrome qubits ($\textmd{Z}$-type or $\textmd{X}$-type), resulting
in four different patterns for the sequence of two-qubit gates in the QEC cycle. Hence, for QEC based on two-qubit gates with driven
couplers, besides shared lines for qubit addressing, an extra four sets of shared lines are required for
selectively choosing gate patterns.}
\label{fig2}
\end{figure*}

For controlling two-dimensional (2D) square qubit lattices, Figure~\ref{fig1}(c) schematically shows the multiplexed control
architecture, which comprises two types of shared control lines, row lines and column lines. The qubits are located on
the intersections of the two-type lines, with qubit connected (coupled) to a distinct pair of control
lines, enabling spatial qubit addressing. By simultaneously applying appropriate pulses to the shared control
lines, each qubit can be driven by a unique pair of control pulses through the associated row and
column lines, as depicted in Fig.~\ref{fig1}(c). As we will discuss in the next section, when the pair of control pulses
satisfies certain conditions, single- or two-qubit gate operations can be realized. Thus, this control strategy can
be well suited for performing parallel addressing, allowing simultaneous single- or two-qubit gates
across the qubit lattice.

However, due to the shared nature of the control lines among qubits, selective addressing of individual qubits or specific
subgroups (i.e., for implementing non-structured quantum circuits) without impacting all other idle (inactive) qubits is
challenging~\cite{Tan2024}. For example, as shown in Fig.~\ref{fig1}(c), when one considers addressing the qubit highlighted
in light orange, two pulses are delivered to the
associated row and column lines. Consequently, idle qubits coupled to the same row or column control lines can be driven
by one of the two shared pulses, causing idling qubit errors. Similarly, one can envision a scenario in which fine-tuned
gate pulses are simultaneously applied to all the shared lines. This allows gate operations on select qubits to be executed
successfully. However, the non-selective (inactive) qubits are also driven by pairs of pulses, potentially leading to idle gate
errors. Note that as each qubit is controlled by a unique pair of gate pulses, this allows for tailored design of pulses
to implement distinct gates on individual qubits. For instance, target gates can be applied to the active qubits while
identity operations can be performed on the idle qubits. However, this approach generally results in the introduction
of additional overheads in gate calibration.

In comparison to the independent control, there exist two features in controlling qubits. The first feature involves the
trade-off between multiplexed control and control flexibility. While multiplexed control can reduce control flexibility, it
effectively addresses wiring challenges by enabling parallel control of $N$ qubits with $O(\sqrt{N})$ lines.
The second feature pertains to the requirement for both single- and two-qubit gates to be executed using a pair of
control pulses, thereby supporting row-column addressing. Consequently, two key issues emerge within this architecture: the
impact of reduced control flexibility on the utility of multiplexed control in quantum computing, and the validation of
gate operations that support row-column addressing. In the subsequent discussion, we aim to first tackle the former
one. We will start with a discussion of being agnostic to the physical details of qubits and gate schemes, which
will be elaborated in the next section.

Despite the compromise in control flexibility, the inherent parallelism of the multiplexed control could allow for
executing structured quantum circuits, which consist of layers of parallel single- and two-qubit gates across qubit
systems (note that by using the isomorphous waveform technique
introduced in Ref.~\cite{Han2024}, an arbitrary quantum circuit comprising modular single- and two-qubit gates can be
compiled into structured quantum circuits). Nevertheless, for practicality and utility, the multiplexed control should be tailored
to be compatible with the specific properties of the structured circuits.

Here, we consider the realization of a QEC circuit with the multiplexed control scheme.
Figure~\ref{fig2}(a) shows the typical quantum circuit for stabilizer measurements in surface code using $\textmd{CZ}$ gates~\cite{Acharya2023a},
where the single-qubit gate (including the identity gate) is decomposed into two $\sqrt{\textmd{X}}$ gates and three $\textmd{Z}$ gates~\cite{McKay2017}
(the rationale behind this decomposition will be addressed in the subsequent section). Throughout the QEC cycle, interleaved single- and two-qubit
gates are implemented and at the end of the cycle, syndrome qubits are measured first and then reset while data qubits are idle. To ensure compatibility with the QEC circuit, Figure~\ref{fig2}(b) presents the multiplexed control architecture comprising two sets of shared row-column
lines for independently addressing date qubits and syndrome qubits. This tailored
setting is adopted for two reasons: (i) during the syndrome measurement and reset, dynamical
decoupling comprising a sequence of single-qubit gates is generally applied to data qubits for
suppressing idling errors (currently the predominant error source)~\cite{Acharya2023a}; (ii) in certain two-qubit gate
schemes, only one of the qubits, such as the syndrome qubit shown in Fig.~\ref{fig2}(c), is driven by
control pulses~\cite{Chow2011,Caldwell2018}.

\begin{figure*}[tbp]
\begin{center}
\includegraphics[width=16cm,height=5cm]{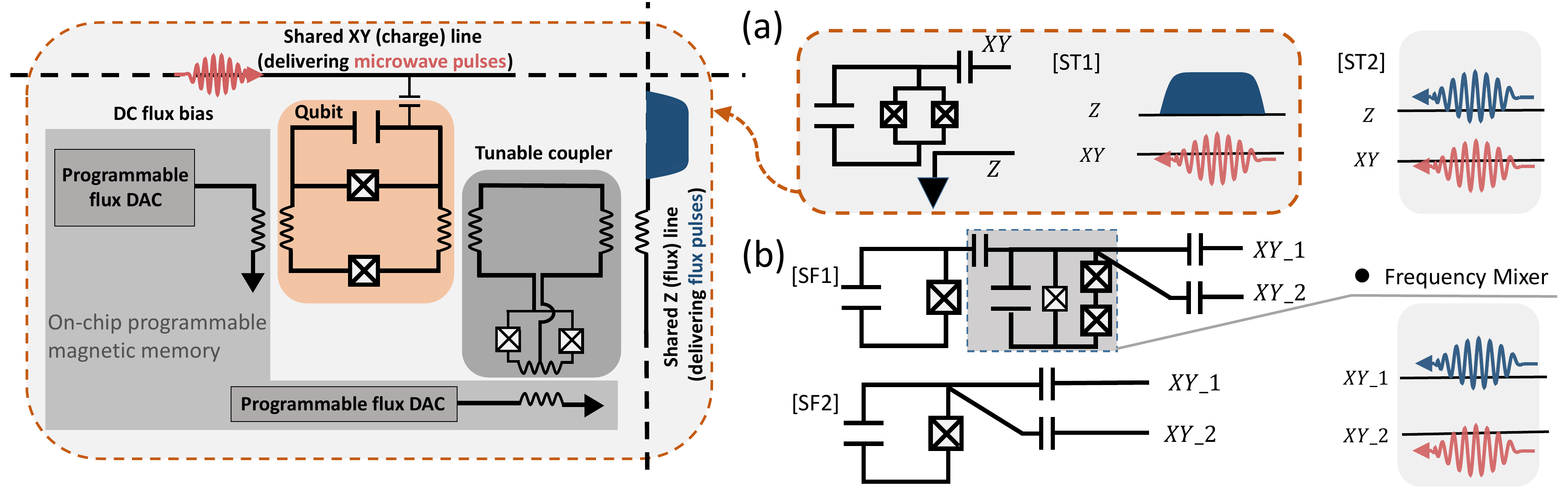}
\end{center}
\caption{Schematic illustration of single-qubit gate schemes with two control pulses.
(a) Single-qubit gates applied to frequency-tunable transmon qubits. During gate operations, a flux pulse and a
microwave pulse are simultaneously delivered to the qubit through the Z line and the XY line, respectively.
Static dc flux biases are needed for biasing the qubits at their idle points and are realized by using an on-chip
programmable magnetic memory such as \emph{$\Phi$}-DAC. (ST1) illustrates the gate scheme based on tuning the qubit on-resonance
with the microwave drive pulse using a baseband flux pulse. Individual tuning of the coupling
between the Z line and the qubit is enabled by a tunable rf-SQUID coupler biased by a \emph{$\Phi$}-DAC. (ST2) shows the gate scheme that
utilizes the combination of parametric drive and microwave drive. Due to the nonlinear dependence
of the qubit frequency on the flux bias, the parametric-driven qubit can have a series of sideband frequency
components. When one of such sidebands is on-resonance with the microwave drive, qubit control can be achieved.
(b) Single-qubit gates for fixed-frequency transmon qubits. During gates, a pair of microwave drive pulses
are simultaneously delivered to the qubit through the two XY lines. (SF1) depicts the gate scheme using the three-wave
mixing process mediated by a flux qubit. When the sum-frequency or difference-frequency of the two drives is on-resonance
with the qubit, the qubit can be controlled by an effective on-resonance drive. (SF2) displays the gate scheme that uses one
off-resonance microwave drive to control a qubit dressed by a second off-resonance drive.}
\label{fig3}
\end{figure*}

Additionally, as depicted in Fig.~\ref{fig2}(c), specific two-qubit gate schemes involve applying control pulses solely to couplers~\cite{McKay2016,Mundada2019,Han2020,Kubo2023}. Meanwhile, in stabilizer measurement circuits of the
surface code, the order of two-qubit gates depends on the syndromes being measured, either $\textmd{Z}$ or $\textmd{X}$. This gives rise
to four different patterns for the sequence of two-qubit gates in QEC cycles~\cite{Terhal2015,Dennis2002}, as shown in Fig.~\ref{fig2}(c).
Therefore, to minimize any aforementioned idle errors and calibration overhead, besides the shared lines for qubit addressing, an
extra four sets of shared lines can be introduced for selectively choosing the gate pattern and addressing couplers.

Before leaving this section, note that although we only focus on QEC circuits, there is potential for adapting
the multiplexed control scheme for logical operations as well. This is supported by the following three
key observations. First, logical operations, such as those based on lattice surgery~\cite{Horsman2012}, in essence, involve different
patterns of stabilizer measurements that are compatible with the multiplexed control. Second, a more tailored
setting can be introduced for patterns of stabilizers that are incompatible with the above setting and the added
line overhead can scale with $\sqrt{N}$. Third, while selectively addressing specific qubit subgroups to accommodate stabilizer
patterns for logical operations may cause qubit idling errors, such as phase errors due to off-resonance drives, this could be
compensated in subsequent QEC cycles, albeit potentially leading to an increase in circuit depth.

\section{Gate operations in the multiplexed control architecture }\label{SecIII}

Here we turn to illustrate how single- and two-qubit gates are realized in the multiplexed control
architecture. As shown in Figs.~\ref{fig1}(c) and~\ref{fig2}, gate operations, unlike that in the
conventional independent control scheme, are implemented by applying simultaneous control pulses to all
shared row and column lines. When focusing on one particular qubit or coupler within the lattice, single-
and two-qubit gate operations are realized by applying a unique pair of control pulses to a qubit or
a coupler. Within this context, we propose various gate schemes that are both compatible with the two-pulse
configuration and the existing superconducting qubit technologies. For illustration purposes, we focus
on the gates applied to transmon qubits~\cite{Koch2007}, but in principle, it should also be feasible for other
superconducting qubits, such as fluxonium qubits~\cite{Manucharyan2009}.

\subsection{Single-qubit gates}\label{SecIIIA}

\begin{table}[ht]
\centering
\caption{Single-qubit gate schemes with two control pulses. The
schematic illustration of these schemes is shown in Fig.~\ref{fig3}. Note that for frequency-tunable
qubits, static dc flux biases (not listed here but shown in Fig.~\ref{fig3}) are generally required to bias the qubits
at their idle points. }
%\begin{ruledtabular}
\begin{tabular}{l|l}
\hline \hline
Specific & Single-qubit addressing strategy
 \\ \hline
\multirow{3}{*}{\shortstack[l]{Tunable \\element}} & \multirow{3}{*}{\shortstack[l]{(ST1) Baseband flux bias + microwave drive~\cite{Bejanin2022,Zhao2023}; \\
\\ (ST2) Flux modulation + microwave drive~\cite{Li2021};}}\\
& \\
& \\ \hline
\multirow{3}{*}{\shortstack[l]{Fixed \\element}} & \multirow{3}{*}{\shortstack[l]{(SF1) Frequency mixer: e.g. three-wave mixing~\cite{Frattini2017,Chapman2023,Liu2014,Zhao2018} ;\\ \\(SF2) Two-tone microwave drive~\cite{Dai2017,He2019,Koong2021,Bracht2021,Wang2023};}} \\
& \\
& \\
\hline \hline
\end{tabular}\label{table1:singleQ}
%\end{ruledtabular}
\end{table}

\begin{figure}[tbp]
\begin{center}
\includegraphics[keepaspectratio=true,width=\columnwidth]{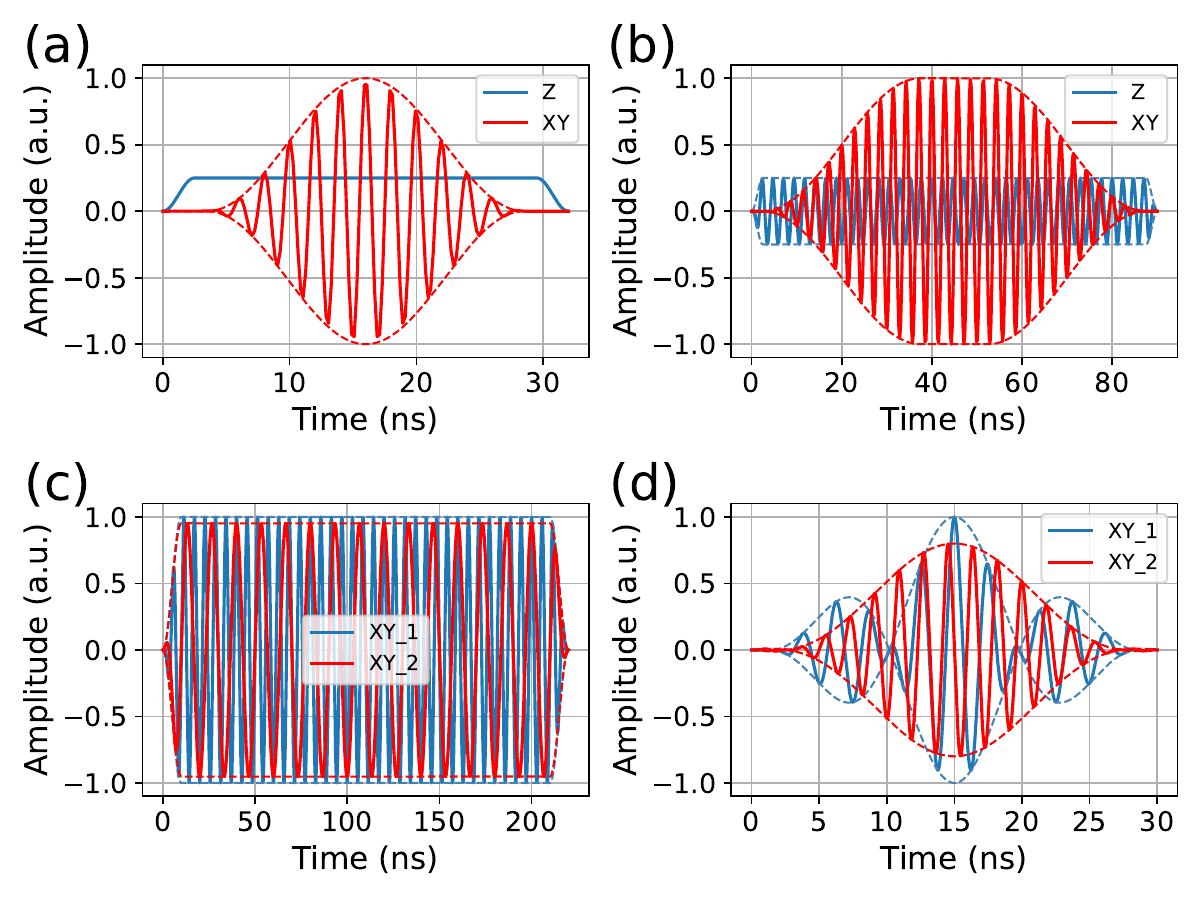}
\end{center}
\caption{The control pulses for single-qubit gate schemes schematically illustrated in Fig.~\ref{fig3}
(also summarized in Table~\ref{table1:singleQ}). The dashed lines denote the pulse envelopes.
(a) shows the pulses for the gate scheme (ST1), using a 25-ns cosine DRAG pulse and a flat-top baseband pulse
with a cosine ramp of 2.5 ns. Here, a delay of 1 ns is inserted between the two pulses, giving rise to
the total gate length of 32 ns. (b) shows the pulses for the gate scheme (ST2), using a flat-top parametric drive
pulse with the ramp time of 2.5 ns and a flat-top microwave drive pulse with the ramp time of 35 ns. The total
gate length is 90 ns. (c) shows the pulses for the gate scheme (SF1), using two 220-ns flat-top microwave drive pulses with
the same ramp time of 10 ns. (d) shows the pulses for the gate scheme (SF2), using two 30-ns DRAG pulses. Note here
that the DRAG coefficient is $\alpha=1$ for suppressing leakage and only the in-phase component of the cosine DRAG pulse
is shown for clarity.}
\label{fig4}
\end{figure}

\begin{figure*}[tbp]
\begin{center}
\includegraphics[width=16cm,height=6cm]{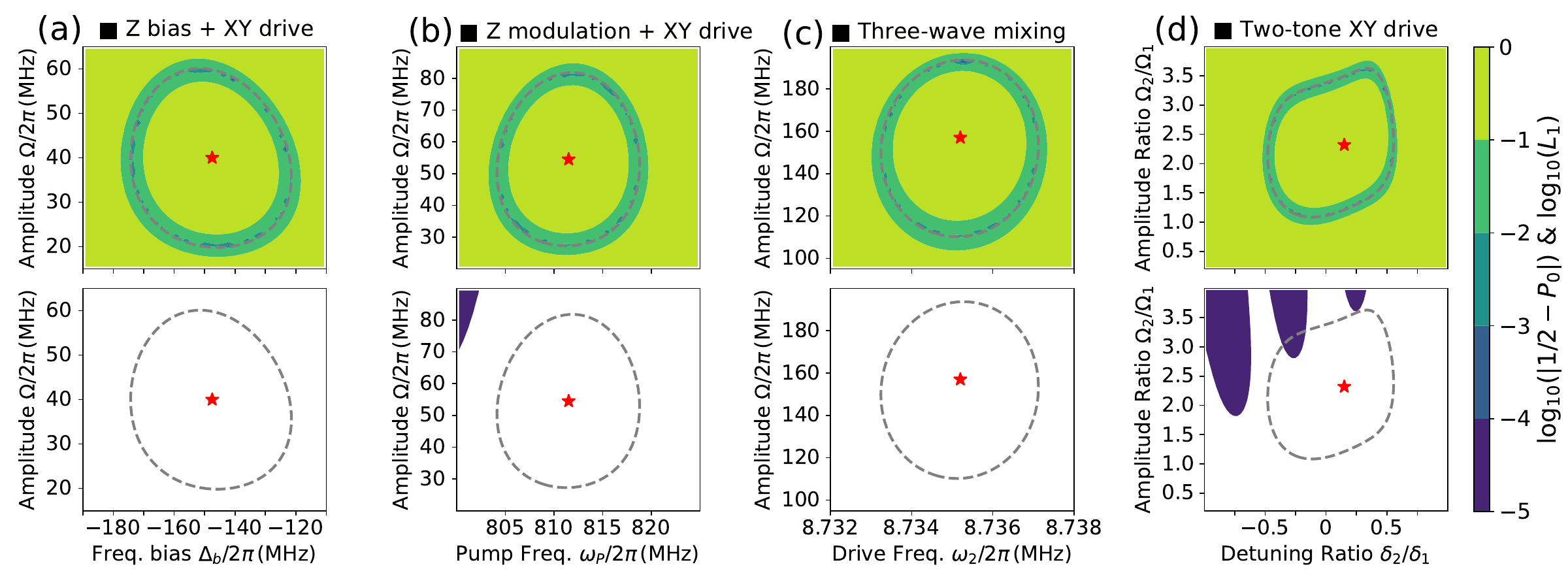}
\end{center}
\caption{The numerical verification of the four single-qubit gate schemes. In the numerical analysis, the
transmon qubit is treated as an anharmonic oscillator~\cite{Koch2007} with the (maximum) frequency
of $\omega_{q}/2\pi=5.25\,\rm GHz$ and the anharmonicity of $\eta_{q}/2\pi=-250\,\rm MHz$. The upper panel shows the
population $P_{0}$ at the end of the applied gate pulses depicted in Fig.~\ref{fig4}, with the qubit prepared in state $|0\rangle$,
while the bottom panel illustrates the leakage $L_{1}$ (white regions indicate leakage below $10^{-5}$). Dashed grey lines
indicate the available regions for realizing $\sqrt{X}$ gates while the red star is for $X$ gates. (a) $P_{0}$ and
$L_{1}$ versus the peak drive amplitude $\Omega$ of the DRAG pulse (with a frequency of $\omega_{d}/2\pi=5.10\,\rm GHz$) and the frequency
bias $\Delta_{b}=\omega_{q}-\omega_{d}$. (b) $P_{0}$ and $L_{1}$ versus the
parametric drive frequency $\omega_{p}$ and the microwave drive amplitude $\Omega$ with the drive frequency of
$\omega_{d}/2\pi=5.75\,\rm GHz$. Here, the dependence of qubit frequency on the flux bias is approximated
by $\omega(\Phi)=(\omega_{q}-\eta_{q})\sqrt{|\cos(\pi \Phi/\Phi_{0})|}+\eta_{q}$~\cite{Koch2007}, where $\Phi_{0}$ denotes the flux quantum.
The used parametric drive is $\Phi=0.05\Phi_{0}+\Phi_{p}(t)\Phi_{0}\cos(\omega_{p} t+\phi_{0})$ (for simplicity, here $\phi_{0}=0$) with
a peak drive amplitude of $\Phi_{p}=0.2$.
The result shows the first-order sideband transition with $\omega_{p}\sim |\omega_{d}-\omega_{q}|$. (c) $P_{0}$
and $L_{1}$ versus the peak drive amplitudes $\Omega$ of the two pulses and the frequency
of the second drive $\omega_{2}$. The frequency of the first drive is $\omega_{1}/2\pi=3.5\,\rm GHz$. Here, the
flux qubit is modeled as a qutrit with cycle transitions $(\omega_{01}/2\pi= 6.25\,{\rm GHz},\,\omega_{02}/2\pi= 10\,{\rm GHz},\,\omega_{12}/2\pi= 3.75\,\rm GHz)$ and transmon-flux coupling strength $(g_{01}/2\pi=94\,{\rm MHz},\,g_{02}/2\pi=140\,{\rm MHz},\, g_{12}/2\pi=136\,{\rm MHz})$ ~\cite{Zhao2018}. (d) $P_{0}$ and $L_{1}$ versus the drive amplitude ratio $\Omega_{2}/\Omega_{1}$ and
the drive-qubit detuning ration $\delta_{2}/\delta_{1}$ of the two pulses, where $\delta_{1(2)}=\omega_{1(2)}-\omega_{q}$.
Here, for the first drive, the peak drive parameters and the drive-qubit detuning are $\Omega_{1}/2\pi=15\rm MHz$
and $\delta_{1}/2\pi=50\,\rm MHz$, respectively. }
\label{fig5}
\end{figure*}

\begin{figure}[tbp]
\begin{center}
\includegraphics[keepaspectratio=true,width=\columnwidth]{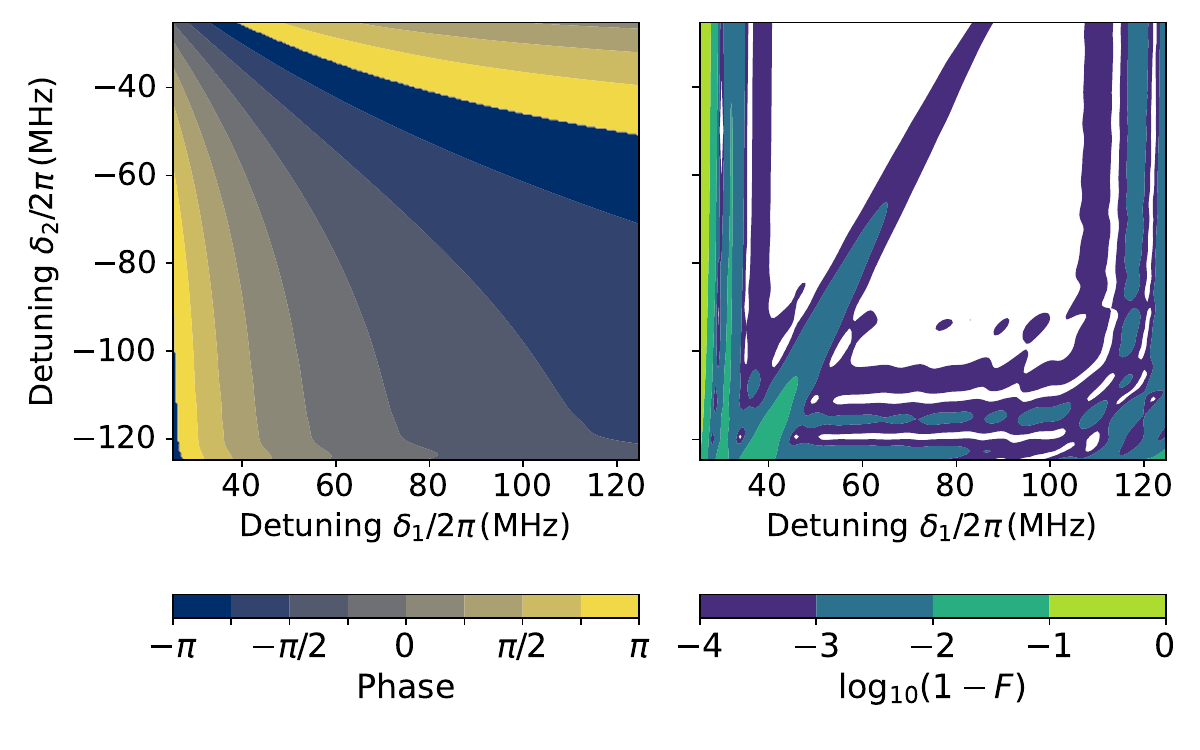}
\end{center}
\caption{Single-qubit Z gates. Z gates can be obtained by utilizing two
off-resonance drives. Similar to Fig.~\ref{fig4}(c), two 150-ns flat-top pulses with
a ramp time of 50 ns are utilized. The left panel shows the angles of the Z rotation versus the two
drive-qubit detunings $\delta_{1}$ and $\delta_{2}$. Meanwhile, the associated Z gate error is shown in the right panel, with white
regions indicating gate errors below $10^{-4}$.}
\label{fig6}
\end{figure}

Table~\ref{table1:singleQ} summarizes four single-qubit gate schemes using two control pulses, which are also
schematically illustrated in Fig.~\ref{fig3}. These schemes can be grouped into two categories, depending on whether
the qubit is tunable. In the following, we will delve into detailed explanations of each of the four
gate schemes.

\emph{Type-ST}: This category pertains to single-qubit gates designed for frequency-tunable qubits.
During gate operations, a flux pulse and a microwave pulse are simultaneously delivered
to the qubit through a shared Z line and a shared XY line, respectively, as shown in Fig.~\ref{fig3}(a).
Furthermore, for frequency-tunable qubits, static dc flux biases are generally needed for biasing the qubits
at their idle points. Following Ref.~\cite{Johnson2010}, this flux offset can in principle be
introduced by using a $\Phi$-DAC, which functions as on-chip programmable magnetic memory and provides
the required static bias. Moreover, the $\Phi$-DAC can be efficiently programmed and addressed by the
addressing circuitry, which requires $O(\sqrt[3]{N})$ control lines for $N$ frequency-tunable qubits~\cite{Bunyk2014}.

(ST1) This control scheme involves a microwave drive pulse and a baseband flux pulse. Similar to Refs.~\cite{Bejanin2022,Zhao2023},
single-qubit gates, such as $\sqrt{X}$ gates, are realized by individually tuning qubits on resonance with the
shared microwave drive. Unlike previous works using an always-on microwave drive~\cite{Zhao2023}, here a microwave drive
pulse is used to prevent any negative effects on other qubit operations, such as qubit readout, qubit reset, and two-qubit
gates. Thus the derivative removal by adiabatic gate (DRAG) can be employed for realizing single-qubit gates~\cite{Motzoi2009}.
Figure~\ref{fig4}(a) shows the typical two-pulse configuration, which consists of a raised cosine flat-top flux pulse for tuning
the qubit frequency from the idle point to the working point and a cosine DRAG pulse for XY control~\cite{Chen2016}. As
qubit idling frequencies should be different from qubits to qubits, the amplitude of the flux pulse applied
to each qubit should be individually fine-tuned to resonate with the shared microwave drive. Given that the flux
pulse is applied globally to qubits through the shared Z lines, individual tuning is required and can be achieved by realizing
tunable coupling between qubits and the shared Z lines. Following Ref.~\cite{Johnson2010}, Figure~\ref{fig3}(a)
shows the tunable rf-SQUID coupler~\cite{Harris2009,Brink2005,Harris2007}, which itself is controlled
by a $\Phi$-DAC, for individually tuning the amplitudes of flux pulses felt by qubits.

(ST2) This scheme entails a parametric drive pulse through the Z line and a microwave drive pulse through
the XY line. As illustrated in Refs.\cite{Li2021,Caldwell2018,Strand2013}, modulating the qubit frequency
with a parametric drive can induce a series of sidebands due to the nonlinear dependence
of the qubit frequency on the flux bias. When one of these sidebands is on-resonance with the microwave drive, it enables coherent
qubit control~\cite{Li2021}. Figure~\ref{fig4}(b) shows this pulse configuration, where a flat-top parametric
drive pulse with cosine-shape ramps modulates the qubit frequency and a raised cosine flat-top microwave drive pulse
facilitates XY control. Note that similar schemes have also been studied recently in semiconductor spin
qubits~\cite{Gyorgy2022,Bosco2023,John2023}.

\emph{Type-SF}: This category pertains to single-qubit gates intended for fixed-frequency qubits.
During single-qubit gates, a pair of microwave drive pulses are simultaneously delivered
to the qubit through the two XY lines, as shown in Fig.~\ref{fig3}(b).

(SF1) Leveraging an on-chip 'three-wave mixer'~\cite{Frattini2017,Chapman2023,Liu2014,Zhao2018}, the two microwave drives
are converted to an effective qubit drive. When the sum-frequency or difference-frequency of the two drives is on-resonance
with the qubit, coherent control of the qubit can be achieved with the effective drive. Figure~\ref{fig3}(b) schematically
illustrates such an on-chip mixer based on flux qubit~\cite{Liu2014,Zhao2018}, which is capacitively coupled to
the transmon qubit and is driven by the two microwave drives. Here, for example, we consider that the
difference frequency of pulses equals the qubit frequency. Accordingly, Figure~\ref{fig4}(c) shows the typical
used raised cosine flat-top pulses.

(SF2) In this scenario, single-qubit gates are realized by applying two off-resonance microwave drives to the
qubit~\cite{Dai2017,He2019,Koong2021,Bracht2021,Wang2023}, as shown in Fig.~\ref{fig3}(b). Such scheme can be understood
as follows: one of the two off-resonance drives is used to dress the (bare) qubit and shift the qubit frequency through
the ac-Stark effect, while the second one is applied for controlling such a microwave-dressed
qubit~\cite{Zhao2022c,Bracht2023,Zuk2023}. Moreover, this two-tone drive scheme can be combined
with the DRAG scheme to suppress leakage errors during gate operations. Figure~\ref{fig4}(d) depicts the typical
used cosine DRAG pulses in this context.

For each of the schemes depicted in Fig.~\ref{fig3} (also summarized in Table~\ref{table1:singleQ}),
Figure~\ref{fig5} presents the numerical verification of their feasibility for performing single-qubit gates, e.g., $\sqrt{X}$
gates or $X$ gates, according to the control pulses shown in Fig.~\ref{fig4}. Here, the upper panel shows the
population ($P_{0}$) in state $|0\rangle$ at the end of the applied pulses with the qubit prepared in its ground
state. The dashed grey line denotes the feasible regions of the control parameters for realizing $\sqrt{X}$ gates while
the red star is for $X$ gates. Furthermore, the bottom panel shows the leakage $L_{1}$~\cite{Wood2018}. These numerical findings
indicate that that low-leakage, high-fidelity single-qubit X rotations can be achieved with the proposed schemes.

As shown in Fig.~\ref{fig2}(a), to achieve universal single-qubit control, we consider compiling arbitrary single-qubit gates
into two $\sqrt{X}$ gates and three physical (or virtual) $Z$ gates~\cite{McKay2017}. The main reason for taking this compiling
method is twofold. First, we prefer the $\sqrt{X}$ gate rather than the $X$ gate due to that the former one can provide more flexibility in choosing
control parameters. This flexibility could provide potential adaptations to address challenges involving inhomogeneities in
coupling efficiencies between qubits and control lines, defects affecting qubit performance~\cite{Zhao2023}, frequency crowding in
choosing control frequencies, and so on. Second, as control
pulses are shared by qubits, individual Z gate applied to each qubit cannot be directly realized by the phase shift of the
shared microwave drive or parametric drive~\cite{McKay2017,Chen2023}, especially in the gate schemes
of (ST1), (ST2), and (SF2). Besides the two main reason, single-qubit errors, that result from microwave crosstalk, could be mitigated by optimizing the Z gates~\cite{Wang2022,Wang2022a} in this decomposition.

\begin{figure*}[tbp]
\begin{center}
\includegraphics[width=12cm,height=7cm]{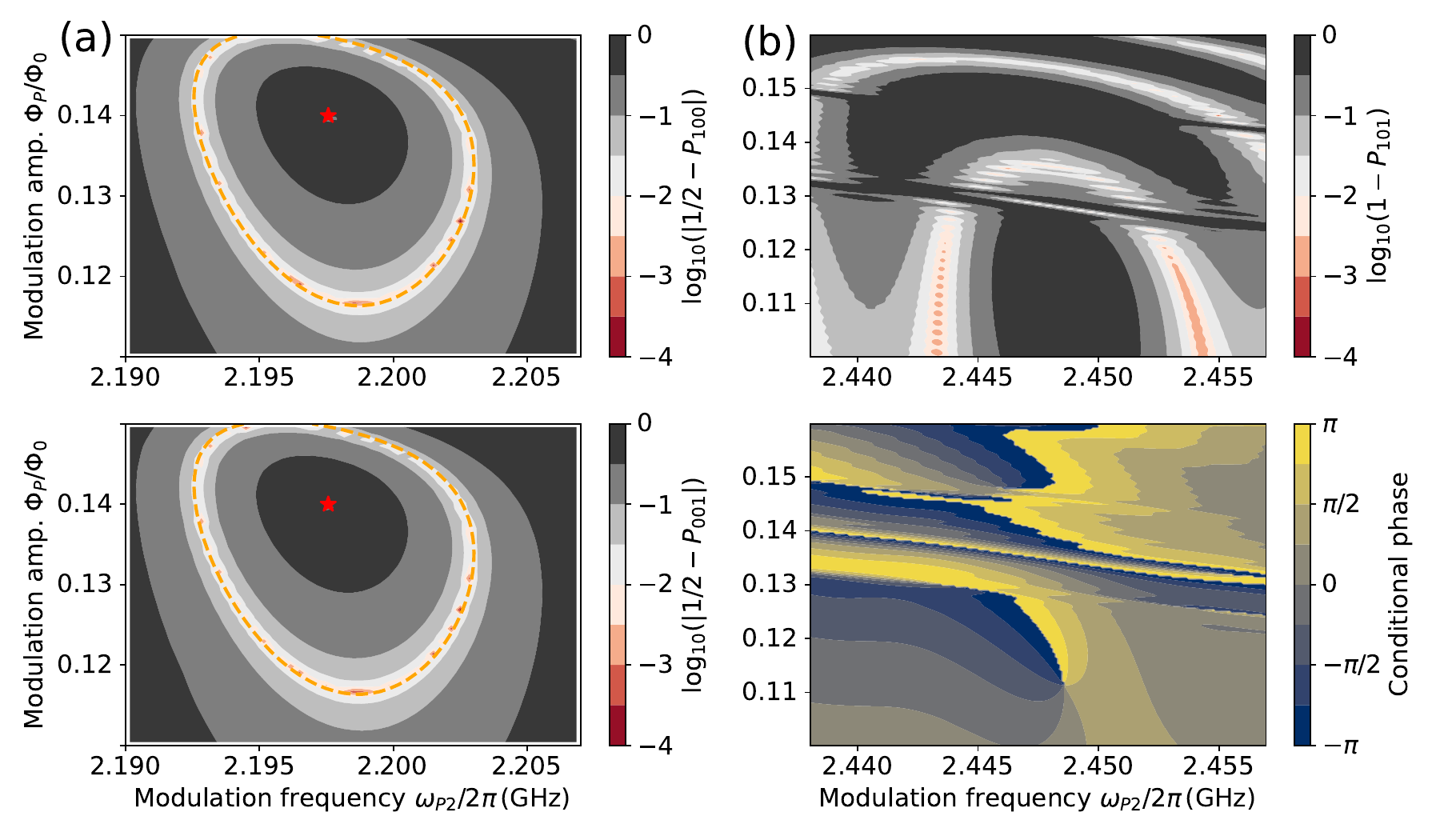}
\end{center}
\caption{The numerical verification of the two-qubit gate operations based on two-tone flux modulations. In the numerical
analysis, we consider that two transmon qubits, labeled by $Q_{a}$ and $Q_{b}$, are coupled via a tunable
bus, labeled by $Q_{bus}$, and the system state is denoted by $|Q_{a}Q_{bus}Q_{b}\rangle$.
The qubit frequencies are $\omega_{a(b)}/2\pi=5.05(5.25)\,\rm GHz$ and the qubit anharmonicity is $\eta_{a(b)}/2\pi=-250\,\rm MHz$. The maximum frequency
and the anharmonicity of the tunable bus are $\omega_{bus}/2\pi=6.20\,\rm GHz$ and $\eta_{bus}/2\pi=-200\,\rm MHz$, respectively. The qubit-bus
coupling strength is $g/2\pi=100\,\rm MHz$.  The dependence of bus frequency on the two-tone flux modulation is approximated by $\omega_{bus}(\Phi)=(\omega_{bus}-\eta_{bus})\sqrt{|\cos(\pi \Phi(t)/\Phi_{0})|}+\eta_{bus}$
with $\Phi(t)=\Phi_{p}(t)[\sin(\omega_{p1}t)+\sin(\omega_{p2}t+\phi_{0})]$ (where $\phi_{0}=0$ for simplicity) and the
modulation frequency $\omega_{p1}/2\pi=2.0\,\rm GHz$. Here, similar to Fig.~\ref{fig4}(c), two raised cosine flat-top pulses
are used and the peak modulation amplitude
is $\Phi_{p}$. (a) Population $P_{100}$ (upper panel) and $P_{001}$ (bottom panel) versus the modulation amplitude $\Phi_{p}$ and
the modulation frequency $\omega_{p2}$ with the qubit prepared in state $|100\rangle$. The ramp time is 15 ns and the total
gate length is 100 ns. Dashed orange lines indicate the available regions for realizing $\sqrt{\textmd{iSWAP}}$ gates while the
red star is for $\textmd{iSWAP}$ gates. Here, the two-qubit gate is actuated when the difference frequency of
the two modulations matches the qubit-qubit detuning. (b) Population $P_{101}$ (upper panel) versus the modulation amplitude $\Phi_{p}$ and the
modulation frequency $\omega_{p2}$ with the system prepared in state $|101\rangle$. The bottom panel shows the associated
conditional phase. The ramp time is 25 ns and the total gate length is 200 ns. Here, the $\textmd{CZ}$ gate is realized
when the difference frequency of the two modulations matches the detuning between the energy levels of $|101\rangle$
and $|200\rangle$. Note that discontinuities near the horizontal line at $\Phi_{p}/\Phi_{0}=0.13$ arise from
the parasitic interaction between the energy levels of $|101\rangle$ and $|110\rangle$.}
\label{fig7}
\end{figure*}

For the gate scheme of (SF1) based on the tree-wave mixing process, the phase of the effective drive applied to the qubit can
be controlled by the difference phase or sum phase of the two pulses. Thus, in principle, the virtual Z gates can be utilized
here. However, as mentioned above, the gate schemes of (ST1), (ST2), and (SF2) necessitate physical Z gates. Generally, this
can be realized by using the ac-Stark effect due to off-resonance drives~\cite{Wei2023}. Here, we consider the implementation
of arbitrary Z gates within (SF2), where two off-resonance microwave drive pulses are applied to the qubit. Similar to
Fig.~\ref{fig4}(c), a pair of raised cosine flat-top pulses are employed for performing Z rotations. Figures~\ref{fig6}(a)
and~\ref{fig6}(b) show the Z rotation angle and the gate fidelity~\cite{Pedersen2007} versus the two drive-qubit
detunings, respectively, illustrating that arbitrary Z rotations can be realized with high fidelity. Note here that
while similar to the gate scheme of (F1) the phase difference can be controlled, the nonlinear dependence of
the qubit dynamics on the frequency difference or phase difference makes it incompatible with the virtual
Z scheme~\cite{Dai2017}, see Appendix~\ref{A} for details.

\subsection{Two-qubit gates}\label{SecIIIB}

\begin{table}[ht]
\centering
\caption{Two-qubit schemes with two control pulses. Similar to the qubit addressing, note that for two-qubit gates
based on frequency-tunable coupler or bus, static dc flux biases (not listed here) are generally needed for
biasing the coupler or bus at their idle points. This can be achieved by using an on-chip programmable magnetic memory, as
shown in the left inset of Fig.~\ref{fig3}(a).}
%\begin{ruledtabular}
\begin{tabular}{l|l}
\hline \hline
Specific & Two-qubit addressing strategy
 \\ \hline
\multirow{3}{*}{\shortstack[l]{Tunable \\element}} & \multirow{3}{*}{\shortstack[l]{(DT1) Baseband flux pulse + local memory~\cite{Zhao2023}; \\
\\ (DT2) Two-tone flux modulation;}}\\
& \\
& \\ \hline
\multirow{2}{*}{\shortstack[l]{Fixed \\element}} & \multirow{2}{*}{\shortstack[l]{(DF1) Frequency mixer: e.g. three-wave mixing~\cite{Frattini2017,Chapman2023,Liu2014,Zhao2018};}} \\
& \\
\hline \hline
\end{tabular}\label{table2:twoQ}
%\end{ruledtabular}
\end{table}

As discussed in Sec.~\ref{SecII}, to be compatible with the multiplexed control architecture, we consider two-qubit gates
based on driving single qubit or a coupler in two-coupled qubit systems. Table~\ref{table2:twoQ} summarizes three possible
two-qubit gate schemes supporting the row-column addressing:

(DT1) For two fixed-frequency qubits coupled via a coupler (bus), two-qubit gates can be
obtained by only tuning the coupler (bus) frequency~\cite{Collodo2020,Xu2020,Zhao2022b,Goto2022,Campbell2023}.
Given the shared flux pulses, each coupler (bus) can be individually tuned by using the on-chip programmable
magnetic memory~\cite{Johnson2010,Bunyk2014,Zhao2023}, as shown in Fig.~\ref{fig3}(a). Note that with the help of
local memory, in principle, only Z lines shared by the couplers or buses are needed to perform parallel two-qubit gates.

(DT2) By leveraging the nonlinear dependence of the transmon qubit or transmon coupler frequency on the flux bias, two-qubit gates
can be realized by applying two-tone flux modulations on the qubits or the coupler. When the the sum or difference frequency
of the two modulations match the subharmonic of the qubit-qubit detuning, two-qubit gates can be
actuated. For example, considering two fixed-frequency transmon qubits coupled via tunable bus~\cite{McKay2016},
Figure~\ref{fig7} shows that two-qubit gates, such as $\textmd{iSWAP}$, $\sqrt{\textmd{iSWAP}}$, and $\textmd{CZ}$ gates
can be obtained by the two-tone modulation of the bus frequency. This two-tone modulation scheme is
adaptable to other coupler circuits~\cite{Mundada2019,Han2020,Kubo2023}. In Appendix~\ref{B}, we further show that
two-qubit gates can be obtained by applying the two-tone flux modulation to one of the coupled
transmon qubits.

(DF1) For two-qubit gates based on driving a single qubit, the three-wave mixer could be used to generate the
desired effective single-qubit drive. However, considering the existing techniques~\cite{Frattini2017,Chapman2023,Liu2014,Zhao2018}, the
strength of the effective drive (here the strength is about $\sim1\,\rm MHz$) is too weak to support the successful implementation
of microwave-activated two-qubit gates, such as cross-resonance (CR) gates~\cite{Chow2011}. Therefore, new physical components
that enable the three-wave mixing process should be introduced to generate effective drives with large amplitudes.

Given the above discussions, the scheme of (DT2) emerges as the most viable option among the three schemes, whereas
the scheme of (DF1) poses the greatest challenge, necessitating the demonstration of new physical components.
In addition to the three discussed schemes, similar to the sing-qubit gate scheme of (SF2), two-qubit gates, which are actuated by
applying two-tone microwave drives to a qubit or a coupler, should be potential solutions for supporting
the row-column addressing and are thus worth exploring in future works.

\section{Challenges towards large-scale multiplexed control architectures}\label{SecIV}

In state-of-the-art superconducting quantum processors, several issues, such as distortions
in control signals~\cite{Gustavsson2013,Barends2014}, crosstalk among control signals~\cite{Acharya2023a}, and
temporal fluctuations of qubit parameters and coherence times (due to factors like two-level systems)~\cite{Muller2019},
are not yet to be well addressed for achieving reliable, accurate quantum computing. This becomes even more challenging
within the multiplexed control architecture.

To be more specific, currently, independent control allows
various active approaches to be employed for alleviating the detrimental effects of these issues on qubits. However,
due to the control limitations, these approaches cannot be directly applied to address similar issues within the
multiplexed control architecture. Consequently, rather than merely attempting to calibrate away these
issues, instead, one might turn to fully suppress or eliminate these issues at the qubit chip level. Nevertheless, given
state-of-the-art technologies, significant advancements in this regard is unlikely to be achieved soon. Additionally, at
the level of control pulses, robust quantum control could be explored to provide resilience against these
issues~\cite{Dong2010}, while at the circuit level, gate compilations with adaptations to address these issues can be
developed~\cite{Wang2022,Wang2022a}.

In addition to the aforementioned well-recognized hurdles, there exist two new challenges that are particularly important in
achieving high-fidelity qubit control within the multiplexed control architecture but are rarely involved in the
traditional independent control architecture. In the following, we will discuss the two challenges to be faced when
considering scaling up the multiplexed control architecture.

\subsection{The non-uniformity of qubit parameters}\label{SecIVA}

As illustrated in Sec.~\ref{SecII} and Sec.~\ref{SecIII}, here the gate pulses
are shared by multiple qubits, thus the gate condition for each qubit is distinctively intertwined
with each other. For clarity, we assume that the coupling efficiencies between control lines and qubits
are all the same and focus on the single-qubit gates (similar results can also be obtained for
the two-qubit gates discussed in Sec.~\ref{SecIIIB}). In the single-qubit gate schemes illustrated
in Sec.~\ref{SecIIIA}, with fixed-length drives featuring the same pulse shape, gates
can only be actuated when the gate conditions
\begin{equation}
\begin{aligned}\label{eq1}
F(\Omega_{i},\Omega_{j},\omega_{i},\omega_{j})=\omega_{ij}
\end{aligned}
\end{equation}
are satisfied. Here $\Omega_{i}$ ($\Omega_{i}$) and $\omega_{i}$ ($\omega_{i}$) denote the amplitude and frequency
of the two drive pulses applied to the qubit (with the frequency of $\omega_{ij}$), which is located at the intersection of the $i$th row line and the $j$th column
line. As mentioned in Sec.~\ref{SecIIIA}, in the following, we focus on $\sqrt{\textmd{X}}$ gates, which provide more
flexibility in setting pulse parameters (similarly, for two-qubt gates, one might prefer $\sqrt{\textmd{CZ}}$ gates~\cite{Shi2023}
or $\sqrt{\textmd{iSWAP}}$ gates, see, e.g., Fig.~\ref{fig7}).

Under the condition of Eq.~(\ref{eq1}) and given fixed drive amplitudes, pulse solutions for the scheme, which relies on the resonance
condition $f(\omega_{i},\omega_{j})=\omega_{ij}$ (see, e.g., Fig.~\ref{fig8} in Appendix~\ref{A}), such as (SF2), (ST1), and (ST2), can in principle
exist. This is because in a square qubit lattice comprising $n\times n$ qubits, $2n$ distinct frequencies should suffice
for actuating parallel gate operations. However, even if solutions exist, considering that the resonance conditions are
generally intertwined with the drive amplitudes and the non-uniformity of qubit parameters, such as the coupling efficiency
between qubits and control lines (i.e., leading to the non-uniformity of the drive amplitude), are ubiquitous in
reality, whether and how such solutions can be realized practically in large-scale systems should be open
questions. This issue warrants further exploration in further works. In the following, we also consider
an alternative approach, i.e., reducing the number of conditions, allowing us to explore another extreme
('trivial') situation.

In the context of 2D square qubit lattices depicted in Fig.~\ref{fig2}(a), the utilization of two sets of shared row-column
lines allow us to selectively address neighboring qubits, i.e., data qubits and syndrome qubits. In this
way, the qubit frequency allocation in each diagonal can follow a zigzag pattern, e.g., where date qubits at a frequency
band with a typical value of $\omega_{D}$ and syndrome qubits at a separate frequency band with
a typical value of $\omega_{S}$. Consequently, the gate conditions for data qubits and syndrome qubits are
decoupled to each other, facilitating a separate treatment for each type of qubits. Hence, in the following,
we focus on the data qubits with gate conditions potentially being reduced to a single criterion denoted by
\begin{equation}
\begin{aligned}\label{eq2}
&F(\Omega_{D,i},\Omega_{D,j},\omega_{D,i},\omega_{D,j})=\omega_{D,ij}\simeq\omega_{D},
\end{aligned}
\end{equation}
where the subscript $D$ indicates the gate condition for data qubits. Finding the solution of the above equation is equivalent
to control of an ensemble of qubits, where the qubit frequencies could be different from qubit to qubit
(given state-of-the-art technologies~\cite{Hertzberg2021,Pappas2024}, the non-uniformity of qubit frequency can
be suppressed to the level of $10\,\rm MHz$~\cite{Hertzberg2021}), using uniform control pulses, i.e., the parameters of the
pulse in each row or column are the same.

However, in fact, besides the qubit frequency and the coupling efficiency, the
non-uniformity of qubit parameters also results from other factors, such as the qubit anharmonicity and the phase difference among
control lines. Therefore, considering all these non-uniformity of qubit parameters, whether one can find the
pulse solution of Eq.~(\ref{eq2}) for multiplexed control of large-scale qubit systems crucially hinges on the
uniformity of qubit parameters. Furthermore, if these non-uniformity issues can be effectively addressed, the
multiplexed control can be simplified to a trivial situation, eliminating the need for row-column addressing
and two-pulse control configurations.

As shown in Fig.~\ref{fig3}(a), in principle, on-chip programmable magnetic memory can be used to
mitigate most of the above-discussed non-uniform issues~\cite{Johnson2010,Zhao2023} but
its compatibility with high-fidelity qubit control has not yet been demonstrated. Meanwhile, robust quantum
control could also be explored to optimize the control pulse against these non-uniformity issues~\cite{Dong2010,Hai2022,Shao2024}.
Given state-of-the-art quantum technologies and the above discussions, we expect that solely adopting one of the two types
of gate solutions supporting the gate conditions in Eqs.~(\ref{eq1}) and~(\ref{eq2}) presents a formidable hurdle in
achieving high-fidelity multiplexed qubit control at scale. Therefore, one might choose to take the combination of such
two solutions, thus enabling the adaptations to address various issues, including the non-uniformity issue.

\subsection{Gate calibration}\label{SecIVB}

While individual gate calibrations and benchmarking may still be possible as in the independent control
architecture~\cite{Kelly2018,Klimov2020}, the multiplexed control should lead to a significant degradation
of the efficiency and the performance that can be reached. This is because that the gate calibration procedures
for each qubit are intertwined with each other. Alternatively, one might also consider fine-tuning up gates at the
quantum circuit level, such as stabilize measurement circuits~\cite{Kelly2016}, where the detection events
can be used for informing and guiding the optimization of gate parameters.

Moreover, when calibrating gate parameters tailored to implement distinct gates on individual qubits, additional
overhead is required. As illustrated in Fig.~\ref{fig2}(a), given the $\sqrt{\textmd{X}}$ gates, arbitrary single-qubit
gates, including identity operations, can be directly optimized by tuning up the Z gates. However, as discussed in
Sec.~\ref{SecII}, selectively activating two-qubit gates on qubit subgroups, such as implementing standard two-qubit
gates (e.g., $\textmd{iSWAP}$-type gates and $\textmd{CZ}$ gates) on selective qubits while applying identity operations
to the inactive qubits, presents a nontrivial task for gate tune-up.

These challenges generally stem from the tradeoff between line overhead and control flexibility. In order to compensate for the
degradation in control flexibility, additional gates or gate patterns need to be tuned up, potentially leading to an
increase in circuit depths~\cite{Tan2024}.

\section{conclusion}\label{SecV}

In conclusion, we introduce a multiplexed control architecture for superconducting qubits with shared
row-column lines. This architecture could in principle provide an efficient approach for parallel controlling $N$
qubits with $O(\sqrt{N})$ control lines. We also propose various single- and two-qubit gate
schemes that are both compatible with the row-column qubit addressing scheme and the existing superconducting
qubit technologies. Leveraging the inherent parallelism of this architecture, we show that the multiplexed control
is suitable for the implementation of structured quantum circuits. As an immediate application, we show that the
architecture can be specifically tailored to execute the quantum error correction with surface code.

We envision that a proof-of-concept demonstration of multiplexed qubit control on a small scale could be feasible with the current
technologies and hope that our work could motivate further experimental and
theoretical research in incorporating shared control into scalable quantum information processing
with superconducting qubits.

\begin{acknowledgments}
The author would like to thank Guangming Xue, Jun Li, Kunzhe Dai, Zhikun Han, and Fei Yan for the insightful discussions.
Thanks also go to Peng Xu, Dong Lan, and Haifeng Yu for their generous support.
%Part of this work was done when the author was working at the Beijing Academy of Quantum Information Sciences.
The author also gratefully acknowledges support from the National Natural Science Foundation of
China (Grants No.12204050) and the Beijing Academy of Quantum Information Sciences.
\end{acknowledgments}

\appendix

\begin{figure}[tbp]
\begin{center}
\includegraphics[keepaspectratio=true,width=\columnwidth]{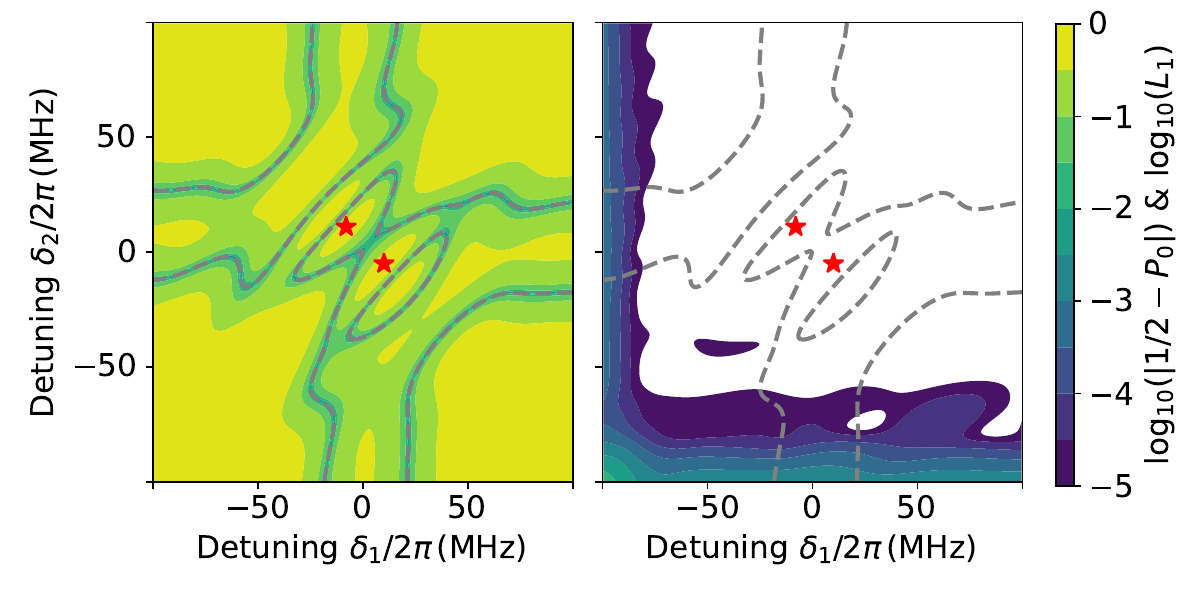}
\end{center}
\caption{Nonlinear dependence on the frequency difference of the two-tone microwave drive.
The left (right) panel shows the population $P_{0}$  (the leakage $L_{1}$) versus the drive-qubit detunings $\delta_{1}$ and $\delta_{2}$ (white regions
are where leakage below $10^{-5}$). Here, the peak drive parameters are $\Omega_{1(2)}/2\pi=25\rm MHz$. Dashed grey
lines indicate the feasible regions for realizing $\sqrt{\textmd{X}}$ gates while the red stars indicate the gate parameters
for implementing $\textmd{X}$ gates.}
\label{fig8}
\end{figure}

\section{Nonlinear dependence on the frequency difference in the two-tone drive scheme}\label{A}

For the two-tone microwave drives of (SF2) studied in Fig.~\ref{fig3}(d), Figure~\ref{fig8} shows
the population $P_{0}$ and the leakage $L_{1}$ versus the drive-qubit
detunings $\delta_{1}$ and $\delta_{2}$, demonstrating the nonlinear dependence of the
single-qubit dynamics on the frequency difference of the two drives. Here, as in Fig.~\ref{fig3}, in the numerical
analysis, the qubit frequency and the anharmonicity are $\omega_{q}/2\pi=5.25\,\rm GHz$
and $\eta_{q}/2\pi=-250\,\rm MHz$, respectively.

\section{iswap gates using two-tone flux modulation of qubits}\label{B}

\begin{figure}[tbp]
\begin{center}
\includegraphics[keepaspectratio=true,width=\columnwidth]{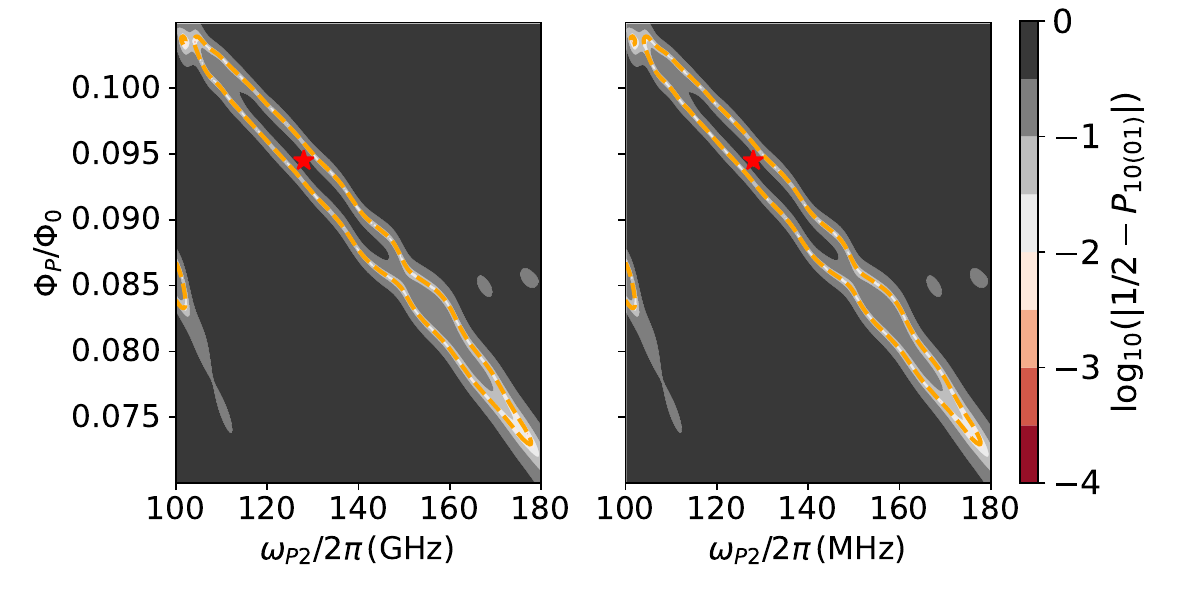}
\end{center}
\caption{The two-qubit gate operations based on applying two-tone flux modulations to qubits. We consider that
two transmon qubits, labeled by $Q_{a}$ and $Q_{b}$, are coupled directly and a two-tone flux modulation is applied
to the frequency-tunable qubit $Q_{a}$. The system state is denoted by $|Q_{a}Q_{b}\rangle$. The left (right) panel
shows population $P_{10}$ ($P_{01}$) versus the modulation amplitude $\Phi_{p}$ and the modulation frequency $\omega_{p2}$
with the system prepared in state $|10\rangle$. Dashed orange lines indicate the available regions for
realizing $\sqrt{\textmd{iSWAP}}$ gates while the red star is for $\textmd{iSWAP}$ gates.}
\label{fig9}
\end{figure}

Here, we consider that one frequency-tunable transmon qubit, labeled by $Q_{a}$, is coupled fixedly to a fixed-frequency
transmon qubit, labeled by $Q_{b}$, with the coupling strength of $g/2\pi=5.5\,\rm MHz$. The qubit frequencies and the
qubit anharmonicities are $\omega_{a(b)}/2\pi=5.05(5.25)\,\rm GHz$ and $\eta_{a(b)}/2\pi-250\,\rm MHz$,
respectively. When applying the two-tone flux modulation to $Q_{a}$, the qubit frequency is approximated
by $\omega_{a}(\Phi)=(\omega_{a}-\eta_{a})\sqrt{|\cos(\pi \Phi(t)/\Phi_{0})|}+\eta_{a}$
with $\Phi(t)=\Phi_{p}(t)[\sin(\omega_{p1}t)+\sin(\omega_{p2}t+\phi_{0})]$. Here, the modulation
frequency is $\omega_{p1}/2\pi=50\,\rm MHz$ and $\phi_{0}=0$ for simplicity. As in Fig.~\ref{fig7}(a),
the raised cosine flat-top pulses are used, with a total gate length of 100 ns and a ramp time of 15 ns. Figure~\ref{fig9}
demonstrates that under such a two-tone flux modulation, two-qubit gates, such as $\sqrt{\textmd{iSWAP}}$ gates
and $\textmd{iSWAP}$ gates, can be actuated.

\end{document}